\documentclass[letterpaper, oneside, 12pt]{article}
\usepackage{amsmath}
\usepackage{graphicx}%
\usepackage{amsfonts}%
\usepackage{amssymb}
\usepackage{xcolor}
\usepackage{overpic}
\usepackage[noend]{algpseudocode}

\usepackage{subcaption}  % preamble에 필요

\usepackage{rotating}
\usepackage{epstopdf}
\usepackage{url}
\usepackage[linesnumbered, ruled, vlined]{algorithm2e}
\usepackage{comment}
\usepackage{lipsum}
\usepackage{enumitem}
\newlist{steps}{enumerate}{1}
\setlist[steps, 1]{label = Step \arabic*:}
\usepackage{mathtools}
\usepackage{amsmath,amsfonts,amsthm,bm} % Math 
\usepackage{multirow}
\usepackage{pgfplots}
\usepackage{natbib}
\usepackage{pst-solides3d}
\usepackage{booktabs}
\usepackage{tabularray}
\usepackage{multirow}
\usepackage[section]{placeins}
\usepackage{subcaption}
\usepackage{newfloat}
\usepackage{caption}

\usepackage{lscape}

\usepackage{setspace}

\usepackage[hmargin=1.0in,vmargin=1.0in]{geometry}

\usepackage[compact]{titlesec}
\titlespacing{\section}{0pt}{*0.8}{*0.8}
\titlespacing{\subsection}{0pt}{*0.8}{*0.8}
\titlespacing{\subsubsection}{0pt}{*0.8}{*0.8}

\title{Deep Probabilistic Spatial Modeling for Multivariate Mixed-Type Responses}

\date{}

\author{
Yeseul Jeon$^{1,2}$, Kyeong Eun Lee$^{3}$, Joon Jin Song$^{4}$\\ 
\small
$^{1}$Department of Statistics, Texas A\&M University, College Station, TX, USA\\
\small
$^{2}$Department of Epidemiology \& Biostatistics, University of California San Francisco, San Francisco, CA, USA\\ 
\small
$^{3}$Department of Statistics, Kyungpook National University, Daegu, Korea\\ 
\small
$^{4}$Department of Department of Statistical Science, Baylor University, Waco, TX, USA. 
}

\newcommand{\bb}{ {\boldsymbol b} }

\newcommand{\boldf}{ {\boldsymbol f} }
\newcommand{\bF}{ {\boldsymbol F} }

\newcommand{\bp}{ {\boldsymbol p} }

\newcommand{\br}{ {\boldsymbol r} }

\newcommand{\bs}{ {\boldsymbol s} }

\newcommand{\bu}{ {\boldsymbol u} }

\newcommand{\bw}{ {\boldsymbol w} }
\newcommand{\bW}{ {\boldsymbol W} }

\newcommand{\by}{ {\boldsymbol y} }

\newcommand{\bz}{ {\boldsymbol z} }

\newcommand{\bphi}{ {\boldsymbol \phi} }
\newcommand{\bPhi}{ {\boldsymbol \Phi} }

\newcommand{\bmu}{ {\boldsymbol \mu} }

\newcommand{\bSigma}{ {\boldsymbol \Sigma} }
\newcommand{\btheta}{ {\boldsymbol \theta} }

\begin{document}

\maketitle

\begin{abstract}
Many scientific applications involve mixed spatially indexed outcomes of heterogeneous types that are driven by shared latent mechanisms. Modeling such data is challenging due to complex, nonlinear, and potentially nonstationary spatial dependence, as well as the need for coherent joint inference across mixed outcome distributions. Existing multivariate mixed outcome models often rely on restrictive linear assumptions, while recent deep learning approaches emphasize predictive flexibility but typically lack coherent joint modeling and uncertainty quantification for spatial data. We develop MultiDeepGP, a scalable and statistically principled framework for joint modeling of multivariate mixed outcomes in spatial settings. The proposed approach introduces a shared latent spatial component that governs cross-outcome dependence while allowing outcome-specific distributions. Spatial dependence and nonlinear structure are captured through a deep latent representation, and uncertainty quantification is enabled via an efficient Monte Carlo–based inference strategy. This construction balances modeling flexibility with probabilistic interpretability and computational feasibility. The proposed method is evaluated through simulation studies designed to reflect key challenges in mixed outcome spatial modeling, as well as an application to georeferenced environmental and public health data from the African Great Lakes region. The results demonstrate that the proposed framework provides accurate joint prediction and reliable uncertainty quantification in complex spatial settings.
\end{abstract}
\noindent\emph{Keywords: Deep Gaussian processes, Joint modeling, Latent spatial processes, Multivariate mixed outcomes, Uncertainty quantification}

\section{Introduction}

In many scientific applications, researchers routinely encounter mixed outcomes of differing types that arise from a common underlying system. For example, environmental and public health studies often involve a combination of binary indicators, count responses, and continuous measurements that reflect complementary aspects of a complex process. Such mixed outcome data are not an exception but rather the norm in real-world settings, particularly when measurements are collected across heterogeneous populations and spatially extended domains.

A common strategy is to model each outcome separately using outcome-specific regression models tailored to the marginal distribution of interest. While this approach is convenient, it ignores the fact that mixed responses are often driven by shared latent mechanisms. In spatial settings, this limitation is particularly acute: environmental drivers, socioeconomic factors, and unobserved spatial processes frequently induce dependence both within and across outcome types. Modeling outcomes independently not only discards information contained in their cross-dependence but can also lead to inefficient estimation, reduced predictive accuracy, and incoherent uncertainty quantification when interest lies in joint prediction or inference.

A substantial literature has therefore focused on joint modeling of multivariate outcomes within the generalized linear mixed model (GLMM) framework. Early work extended GLMMs to accommodate mixed responses from the exponential family by introducing correlated random effects, enabling dependence across outcomes through shared latent components~\citep{gueorguieva2001multivariate}. Subsequent developments have applied multivariate GLMMs to a wide range of applied problems, including biomedical and epidemiological studies, under both frequentist and Bayesian paradigms \citep{jaffa2015multivariate, xu2024causal}. These models provide a principled and interpretable framework for joint modeling, but their practical applicability is often limited by restrictive assumptions on linear predictors, parametric random-effect distributions, and scalability, especially in high-dimensional or spatially structured settings.

An alternative line of research introduces latent variable models to capture dependence among mixed outcome types. By positing one or more unobserved continuous latent processes that link heterogeneous responses, these approaches offer greater flexibility in modeling cross-outcome association while respecting outcome-specific marginal distributions \citep{mcculloch2008joint}. Latent variable models for mixed data have been successfully applied in diverse contexts, including neuroimaging, cognitive assessment, and clinical trials \citep{gruhl2013semiparametric, samani2010multivariate, marchese2018joint, mcmenamin2019sample}. However, most existing formulations rely on low-dimensional latent structures, linear measurement models, and simplified dependence assumptions, which can be inadequate for capturing complex spatial variation and nonstationarity commonly observed in real-world data.

To address the limited flexibility of classical joint models that rely on linear predictors and relatively low-dimensional random-effect structures, a growing body of work has explored deep learning as a scalable alternative for modeling complex multivariate relationships in large datasets. In particular, multi-task learning architectures provide a natural way to learn shared representations across outcomes while allowing outcome-specific heads to respect distinct data-generating mechanisms \citep{caruana1997multitask, zhang2018survey}. Related developments also aim to integrate mixed-effects structure into neural networks to better accommodate clustered or correlated data \citep{simchoni2023integrating, mogharabin2024mcgmenn}. While these approaches offer improved predictive flexibility, they are typically developed for independent or weakly dependent data, and they rarely address the key challenges that arise when the dependence is inherently spatial and the outcomes are heterogeneous.

In spatial applications, ignoring dependence across nearby locations can lead to miscalibrated uncertainty and degraded predictive performance, motivating a recent surge of interest in statistical deep learning for spatial and spatiotemporal data \citep{wikle2023statistical}.

\citet{lin2023enhancements} focus on the coordinate-to-feature mapping, refining how spatial basis functions are constructed and used to embed locations before they enter the network, with the aim of improving stability and predictive accuracy. Separately, \citet{nag2023spatiotemporal} target to model temporal structure, extending basis-embedded networks to handle space-time indexed processes for interpolation and forecasting. On the other hand, DeepKriging~\citet{chen2024deepkriging} focuses on delivering a neural network architecture that mirrors the core structure of kriging-based spatial prediction. The model embeds spatial basis functions of geographic coordinates within a neural network to capture spatial dependence while retaining the computational advantages of stochastic optimization. These contributions demonstrate that embedding-based neural architectures can provide an effective and scalable alternative to classical kriging and cokriging for large spatial datasets. However, extending these approaches to mixed-outcome settings introduces additional challenges. In particular, joint modeling of heterogeneous outcomes requires explicit specification of outcome-specific likelihoods and a principled mechanism for capturing cross-outcome dependence. %These components are not incorporated in current DeepKriging-type models, which are primarily formulated for single or low-dimensional continuous outcomes.

A separate but equally important challenge is uncertainty quantification. Multi-output deep learning models substantially increase the dimensionality and structural complexity of the parameter space, inducing highly coupled dependence structures across outputs and layers. As a result, posterior inference becomes high-dimensional and analytically intractable, making scalable and well-calibrated uncertainty quantification difficult in practice. Consequently, many deep learning models prioritize point accuracy over principled uncertainty assessment, even though scientific decision making often requires calibrated predictive uncertainty, especially in spatial settings where extrapolation and data sparsity are common \citep{wikle2023statistical}. Bayesian deep learning~\citep{neal2012bayesian} provides one principled route to uncertainty quantification by placing probability distributions over network weights. However, fully Bayesian training and inference can be computationally prohibitive at scale~\citep{blei2017variational, salimbeni2017doubly, van2020uncertainty}, and become even more challenging when combining spatial dependence with mixed likelihoods. A particularly attractive compromise is Monte Carlo dropout~\citep{gal2016dropout}, which offers a simple and scalable approximation to Bayesian inference and has a direct interpretation as an approximation to inference in deep Gaussian processes~\citep{damianou2013deep}. This connection motivates approaches that retain the flexibility of deep models while enabling uncertainty quantification with minimal additional computational overhead.

Motivated by these gaps, the goal of this paper is to develop a computationally efficient and statistically principled framework for joint modeling of multivariate mixed outcomes in the presence of spatial dependence. We propose a deep latent spatial model that learns a shared spatial component governing cross-outcome dependence while allowing outcome-specific likelihoods and delivering coherent uncertainty quantification through scalable Monte Carlo-based inference. This enables joint prediction and inference for heterogeneous spatial responses in modern large-scale applications where classical multivariate mixed models and fully Bayesian spatial models can be difficult to deploy.

The remainder of the paper is organized as follows. Section 2 provides the scientific motivation and describes the data source of our study. Section 3 introduces the deep latent spatial structure for joint modeling of multivariate mixed outcomes, along with the associated inference and uncertainty quantification strategy. Section 4 presents two simulation studies designed to isolate key challenges in mixed outcome spatial modeling. Section 5 applies the proposed method to georeferenced environmental and public health data from the African Great Lakes region, assessing predictive performance and uncertainty quantification in a real-world context. Finally, Section 6 concludes with a discussion of limitations, potential extensions, and directions for future research.

\section{Scientific Motivation and Data Description}
\subsection{Environmental--health coupling in the African Great Lakes region}

The African Great Lakes region exhibits pronounced heterogeneity in climate, land cover, and access to natural resources, all of which are known to shape malaria transmission dynamics and population vulnerability~\citep{nicholson2017climate, hay2005urbanization, snow2005malaria, weiss2019mapping}. Environmental factors such as rainfall, surface water availability, and vegetation dynamics play an important role in shaping mosquito habitats and human exposure patterns, and are therefore closely linked to malaria transmission~\citep{hay2005urbanization, tompkins2013vegetation, mordecai2019thermal}. Characterizing how these factors co-vary across space, through their shared latent spatial structure across outcomes, is essential for identifying regions exhibiting jointly elevated intensity and prioritizing structurally vulnerable environments. Although each outcome displays its own spatial heterogeneity, the joint latent representation uncovers regions of concurrent elevation, suggesting the presence of common spatial drivers that would remain obscured under separate analyses. At the same time, these environmental processes exhibit strong spatial structure and marked regional variability driven by topography, land use, and climate regimes~\citep{nicholson2017climate, hay2005urbanization}, further underscoring the need for flexible multivariate spatial modeling.

Especially, vegetation conditions, water availability, and malaria incidence should not be viewed as isolated processes. Vegetation structure influences local microclimates and provides ecological conditions conducive to mosquito breeding and survival, while surface water availability directly affects the formation and persistence of larval habitats. In many parts of sub-Saharan Africa, rainfall-driven vegetation dynamics and hydrological conditions jointly regulate mosquito population density and seasonal transmission intensity, thereby shaping spatial patterns of malaria risk. At the same time, access to water resources is closely tied to human settlement patterns and daily behaviors, which in turn modulate exposure pathways and vulnerability to malaria infection. Consequently, malaria incidence reflects the combined effects of environmental suitability, hydrological conditions, and human–environment interactions, rather than a single isolated driver \citep{githeko2000climate, patz2005impact, mordecai2019thermal}.

We illustrate the proposed framework using georeferenced survey data from the 2015 Demographic and Health Surveys (DHS) \citep{icf2015demographic}. The analytic sample consists of $4,741$ survey clusters distributed across nine contiguous countries in the African Great Lakes region, including Burundi, the Democratic Republic of the Congo, Malawi, Mozambique, Rwanda, Tanzania, Uganda, Zambia, and Zimbabwe. Each cluster is associated with a latitude longitude pair and a set of environmental and health measurements collected over the same time period. Since the relationship between environmental conditions and malaria incidence is spatially heterogeneous and not well captured by models that rely on a single outcome or assume fixed covariate effects~\citep{cressie2011statistics, banerjee2014hierarchical}. Motivated by these considerations, our objective is not to explain malaria incidence through environmental predictors alone, but to jointly characterize the spatial configuration of environmental conditions and health outcomes. By modeling vegetation, water availability, and malaria incidence as coupled spatial responses, we aim to capture their shared latent spatial structure, thereby revealing coherent patterns of concurrent elevation that may reflect common underlying drivers. In addition, we quantify predictive uncertainty for each outcome, allowing the model to identify regions where estimates are less certain, particularly in areas with sparse or uneven surveillance data.

This data setting presents several challenges for spatial statistical modeling. First, the response vector comprises variables of mixed data types, including binary, count, and continuous outcomes, which precludes the use of standard Gaussian based multivariate spatial models. Second, the data exhibit strong spatial dependence over a large and irregular geographic domain, with survey locations that are unevenly distributed across countries and ecological zones. Third, the environmental and health outcomes are expected to share latent spatial structure driven by common environmental processes, yet also display outcome specific variation. Finally, our inferential goals extend beyond marginal prediction of a single outcome. Therefore, we seek to jointly predict environmental and health variables while providing calibrated uncertainty quantification that accounts for cross outcome dependence. These considerations motivate the coupled spatial modeling framework introduced in the next section, which is designed to accommodate mixed outcome types, shared spatial structure, and scalable inference over large spatial domains.

\section{Model}

Let $\{\bs_i\}_{i=1}^N \in \mathbb R^{N \times d}$ denote spatial locations and let $\by(\bs_i) = \big(\by_1(\bs_i), \ldots, \by_J(\bs_i)\big)^\top$ denote a vector of heterogeneous responses observed at location $\bs_i$. Given the same location $\bs$, dependence across outcomes is naturally induced through shared spatial structure.  To model this dependence, we introduce a shared latent spatial process $H(\bs_i)$ defined over space. This latent process captures common spatial variation across outcomes, while allowing outcome-specific components through separate likelihoods. Conditioning on a shared latent process $F(\bs_i)$, we assume conditional
independence across outcomes,
\begin{equation*}
p\big(\by(\bs_i) \mid H(\bs_i)\big)
= \prod_{j=1}^J p\big(\by_j(\bs_i) \mid H(\bs_i) \big).
\end{equation*}

\subsection{Nonlinear shared representation using neural networks}
To accommodate complex nonlinear and potentially nonstationary spatial relationships in a computationally scalable manner, we parameterize the shared latent process using a deep neural network. Multilayer neural architectures provide a flexible function class capable of representing highly nonlinear transformations, while allowing efficient optimization in large-scale settings. Specifically, we construct a shared deep feature map by stacking $L-1$ neural network layers, with the $\ell$th layer comprising $k_\ell$ hidden units,
\begin{align}
\bphi_{i,1}  
&= \sigma_1\!\big(\bW_{1} \bs_i + \bb_{1}\big)\odot \br_{1}, \nonumber\\
\bphi_{i,\ell} 
&= \sigma_{\ell}\!\big(\bW_{\ell} \bphi_{i,\ell-1} + \bb_{\ell}\big)\odot \br_{\ell}, 
\qquad \ell=2,\ldots,L-1,
\label{eq:shared-nn}
\end{align}
where $\sigma_\ell(\cdot)$ denotes the activation function at layer $\ell$, $\bW_{\ell}\in\mathbb{R}^{k_\ell\times k_{\ell-1}}$ and $\bb_{\ell}\in\mathbb{R}^{k_\ell}$ are layer-specific weight matrices and bias vectors, and $\odot$ denotes element-wise multiplication. The vectors $\br_{\ell}\in\mathbb{R}^{k_\ell}$ are elementwise multiplicative dropout masks, with entries drawn independently as $r_{\ell,j}\sim\mathrm{Bernoulli}(p_\ell)$ for $j=1,\ldots,k_\ell$, where the dropout probability satisfies $p_\ell\in[0,1]$, to mitigate overfitting. The resulting vector $\bphi_{i,\ell} \in \mathbb{R}^{k_\ell}$ represents the nonlinear transformation feature at $\ell$th layer. The shared nonlinear spatial representation is defined as $H(\bs_i) = \bphi_{i,L-1}$. For each outcome $j=1,\ldots,J$, we introduce an outcome-specific mapping that links the shared representation to the natural parameter,
\begin{equation}
\eta_{j}(\bs_i) = \sigma^{(j)}_L(\bW^{(j)}_{L} \bphi_{i,L-1} + b^{(j)}_{L}), \quad j= 1,\ldots, J, 
\label{eq:outcome-head}
\end{equation}
where $\bW^{(j)}_L\in\mathbb{R}^{1 \times k_{L-1}}$ and $b^{(j)}_L\in\mathbb{R}$ outcome-specific weight and bias parameters. The function $\sigma^{(j)}_{L}(\cdot)$ serves as an outcome-specific link, allowing different distributional forms across responses. Depending on the type of outcome, $\sigma^{(j)}_{L}(\cdot)$ is chosen accordingly; for example, it may be the identity link for continuous Gaussian responses, the logistic link for binary outcomes, or the log link for count data. Conditional on the shared representation, we assume independence across outcomes. Consequently, the likelihood factorizes as
\begin{equation}
p\big(\by(\bs_i)\mid H(\bs_i)\big)
= \prod_{j=1}^J p\big(y_j(\bs_i)\mid \eta_j(\bs_i)\big). 
\label{eq:joint-nn}\end{equation} 

Let $\btheta = \{ \{\bW_\ell, \bb_\ell\}_{\ell=1}^{L-1}, \{ \bW^{(j)}_L, \bb^{(j)}_L \}_{j=1}^J\}$ denote the collection of all network parameters. The network parameters are estimated by minimizing the summed negative log-likelihood across locations and outcomes,
\begin{equation}
\mathcal{L}(\btheta)
= - \sum_{i=1}^N \sum_{j=1}^J \log p\!\left(y_{j}(\bs_i)\mid \eta_{j}(\bs_i)\right)
\;+\; \sum_{l=1}^L \lambda_\ell  \|\btheta_\ell\|_F^2,
\label{eq:nn-loss}
\end{equation}
where $\lambda_\ell$ are penalty parameters controlling shrinkage for weight and bias parameters. The neural network parameterization in Eq.\eqref{eq:joint-nn} provides a scalable and flexible representation of nonlinear spatial structure. However, as formulated, the mapping from $\bs_i$ to the shared representation $H(\bs_i)$ is deterministic given the network parameters. Consequently, this construction alone does not yield a full probabilistic specification of the latent spatial process and does not directly provide uncertainty quantification for predictions.

\subsection{Deep Gaussian Process Models with Mixed Likelihoods via Neural Networks}

To recover a principled probabilistic interpretation, we view the neural network representation in Eq.\eqref{eq:shared-nn} as a finite-dimensional approximation to a deep Gaussian process. Let $\boldf_{i,1}=\bW_1\bs_i + \bb_1 \in \mathbb{R}^{k_1}$ and for $\ell = 2, \ldots, L-1$, $\boldf_{i,\ell} = \bW_\ell \bphi_{\ell-1}+\bb_{\ell} \in \mathbb{R}^{ k_{\ell}}$. We denote by $\bF_{\ell}=[\boldf_{1,\ell}, 
\ldots, \boldf_{N,\ell}]\in \mathbb{R}^{N\times k_\ell}$ the matrix of linear features at layer $\ell$, and $\bPhi_\ell = [\bphi_{1,\ell},\ldots,\bphi_{N,\ell}]$, the corresponding nonlinear transformations. Assigning independent Gaussian priors to the weights and biases, $\bW_\ell \sim p(\bW_\ell)$ and $\bb_\ell \sim p(\bb_\ell)$, the $k$th node ($k=1,\ldots,k_\ell$) in $\ell$th layer admits a Gaussian process representation conditional on the previous layer:  
\begin{equation}
    \bF^{(k)}_\ell |\bF_{\ell-1} \sim N(\mathbf{0}, \bSigma_\ell), \quad \ell = 2,\ldots,L-1,
\end{equation}
where the covariance $\bSigma_\ell \in \mathbb R ^{N \times N}$ is given by
\begin{equation}
    {\bSigma}_{\ell} = \frac{1}{k_{\ell}}\sigma_l(\bPhi_{\ell-1}\mathbf{W}^{\top}_{\ell}+ \mathbf{1}\mathbf{b}_{\ell})
    \sigma_\ell(\bPhi_{\ell-1}\mathbf{W}^{\top}_{\ell}+ \mathbf{1}\mathbf{b}_{\ell})^{\top}.
    \label{covfuncsmultilayer}
\end{equation}
As the layer width $k_l$ increases, the covariance in Eq.~\eqref{covfuncsmultilayer} provides an increasingly accurate approximation to the true Gaussian process kernel~\citep{gal2016dropout}.

\begin{figure}[htbp]
\begin{center}
\includegraphics[width = 0.7\textwidth]{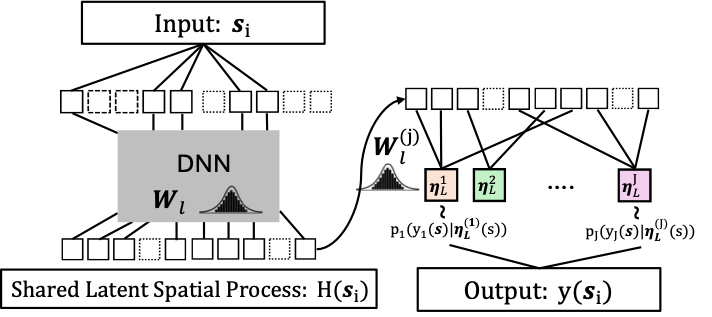}
\end{center}
\caption[]{Illustration of the proposed MultiDeepGP framework. Conditional on the shared latent representation, outcomes are assumed independent, while dependence across responses is induced through the common deep latent spatial process.
}
\label{Fig:model}
\end{figure}

At the final layer, we introduce outcome-specific mappings. For each outcome $j=1,\ldots,J$, define $\boldf^{(j)}_{i,L} = \bW^{(j)}_{L}\bphi_{i,L-1}+\bb^{(j)}_L$ and  $\bF^{(j)}_{L}= [\boldf^{(j)}_{1,L}, 
\ldots, \boldf^{(j)}_{N,L}]\in \mathbb{R}^{N\times 1}$. Conditional on the shared representation $\bF_{L-1}$, $\bF^{(j)}_{L}$ admits a Gaussian process representation:
\[
\bF^{(j)}_L | \bF_{L-1} \sim N(0,\bSigma^{(j)}_L),
\
\]
where covariance matrix $\bSigma^{(j)}_L \in \mathbb R ^{N \times N}$ is 
\begin{equation*}
    {\bSigma}^{(j)}_{L} = \frac{1}{k_{L}}\sigma^{(j)}_L(\bPhi_{L-1}\mathbf{W}^{(j)\top}_{L}+ \mathbf{1}\mathbf{b}^{(j)}_{L})
    \sigma^{(j)}_L(\bPhi_{L-1}\mathbf{W}^{(j)\top}_{L}+ \mathbf{1}\mathbf{b}^{(j)}_{L})^{\top}.
    \label{covfuncsmultilayer2}
\end{equation*}
Note that the $\sigma^{(j)}_L$ determines the outcome-specific link, thereby allowing different response types across outcomes. The corresponding natural parameter is $\boldsymbol{\eta}^{(j)}_{L} = \sigma^{(j)}_L (\bPhi_{L-1}\mathbf{W}^{(j)\top}_{L}+ \mathbf{b}^{(j)}_{L})$. Assuming $k_{L}=1$ for each outcome, the resulting multi-output deep Gaussian process model with mixed likelihoods is defined hierarchically as
\begin{equation}
  y_{j}(\bs)| \bF^{(j)}_{L}(\bs) \sim p_j(y_j(\bs)| {\eta}^{(j)}_{L}(\bs)), 
\label{eq:deepGpmodel}
\end{equation}
where $p_j(\cdot)$ denotes an outcome-specific likelihood from the exponential family. Fig.~\ref{Fig:model} presents a schematic overview of the proposed MultiDeepGP framework.

\subsection{Inference}
Since fully Bayesian inference for deep Gaussian processes via Markov chain Monte Carlo (MCMC) is computationally demanding and difficult to implement in practice, we instead approximate the deep GP in Eq.~\eqref{eq:deepGpmodel} using MC dropout. Specifically, following \citet{gal2016bayesian}, dropout training can be interpreted as performing approximate variational inference in a deep Gaussian process model, where Monte Carlo sampling with dropout masks provides a stochastic approximation to the predictive distribution. 

Let independent standard normal priors are assigned to all weights and biases, i.e., $p(\btheta)=\mathcal N(\mathbf 0,\mathbf I)$, and the intractable posterior over network parameters is approximated using a factorized variational family. A variational distribution $q(\btheta)$ is specified to factorize across layers and outcomes,
\[
q(\btheta) = \prod_{\ell=1}^{L-1}q(\bW_\ell)q(\bb_\ell)\prod_{j=1}^{J}q(\bW^{(j)}_L)q(\bb^{(j)}_L).
\]
For each shared layer $\ell=1,\ldots,L-1$, let $\bw_{\ell,a}\in\mathbb R^{k_{\ell-1}}$ denote the $a$th row of $\bW_\ell$, and let $b_{\ell,a}$ be the corresponding bias.
The variational distribution is defined by row-wise factorization with a two-component Gaussian mixture,
\begin{align*}\label{eq:gal_node_dropout_shared}
q(\bW_\ell,\bb_\ell) &=\prod_{a=1}^{k_\ell} q(\bw_{\ell,a}, b_{\ell,a}),  \nonumber \\ 
q(\bw_{\ell,a}, b_{\ell,a})
=
p_\ell\,\mathcal N\!\Big( [\bmu^w_{\ell,a}, \mu^b_{\ell,a}]^\top,\, &\sigma^2 \mathbf I \Big)
+
(1-p_\ell)\,\mathcal N\!\Big(\mathbf 0,\, \sigma^2 \mathbf I\Big),
\quad p_\ell\in[0,1],
\end{align*}
where $\bmu^w_{\ell,a}\in\mathbb R^{k_{\ell-1}}$ and $\mu^b_{\ell,a}\in\mathbb R$ are variational mean parameters and $\sigma^2$ controls the dispersion.
This specification induces node dropout since the entire incoming weight vector and bias of unit $a$ are simultaneously shrunk toward zero with probability $(1-p_\ell)$. For the output layer of outcome $j$, the same construction is applied as 
\begin{align*}
q(\bW_L^{(j)},\bb_L^{(j)})&=\prod_{a=1}^{k_L^{(j)}} q(\bw_{L,a}^{(j)},  b_{L,a}^{(j)}), \\
q(\bw_{L,a}^{(j)}, b_{L,a}^{(j)})=
p_L^{(j)}\,\mathcal N\!\Big( [\bmu^{w,(j)}_{L,a}, \mu^{b,(j)}_{L,a}]^\top,\, &\sigma^2 \mathbf I \Big)
+ (1-p_L^{(j)})\,\mathcal N\!\Big(\mathbf 0,\, \sigma^2 \mathbf I\Big),\quad p_L^{(j)}\in[0,1], 
\end{align*}
where $\bmu^{w,(j)}_{L,a}\in\mathbb R^{k_{L-1}}$ and $\mu^{b,(j)}_{L,a}\in\mathbb R$ are outcome-specific variational mean parameters for the incoming weight vector and bias of unit $a$ in the $j$th output head, respectively, $\sigma^2$ controls the dispersion of each Gaussian component, and $p_L^{(j)}\in[0,1]$ denotes the inclusion probability for the $j$th output layer. Smaller $p_L^{(j)}$ increases the probability of shrinking $(\bw^{(j)}_{L,a},b^{(j)}_{L,a})$ toward zero, corresponding to node dropout in the $j$th head.

Variational parameters are learned by maximizing an evidence lower bound (ELBO), equivalently minimizing the Kullback--Leibler divergence between the variational approximation $q(\btheta)$ and the exact posterior $p(\btheta\mid\mathcal D)$. The ELBO is defined as
\begin{equation*}\label{eq:elbo_def_new}
\mathcal L_{\mathrm{VI}}(q)
=
\mathbb E_{q(\btheta)}\!\left[\log p(\mathcal D\mid\btheta)\right]
-
\mathrm{KL}\!\left(q(\btheta)\,\|\,p(\btheta)\right).
\end{equation*}
Conditional on $\btheta$, outcomes are assumed independent across locations and outcome indices, so that
\begin{equation*}\label{eq:lik_factor_new}
p(\mathcal D\mid\btheta)
=
\prod_{i=1}^{N}\prod_{j=1}^{J}
p_j\!\left(y_j(\bs_i)\mid {\eta}^{(j)}_L(\bs_i;\btheta)\right),
\end{equation*}
where $p_j(\cdot\mid\eta)$ denotes the exponential-family (or outcome-specific) likelihood for outcome $j$,
and $\boldsymbol{\eta}^{(j)}_L(\bs_i;\btheta)$ denotes the output of the final layer of the deep GP, obtained by applying the last-layer affine map to the learned deep feature representation and then the outcome-specific link transformation. The expected log-likelihood term becomes
\begin{equation}\label{eq:elbo_expand_new}
\mathbb E_{q(\btheta)}\!\left[\log p(\mathcal D\mid\btheta)\right]
=
\sum_{i=1}^{N}\sum_{j=1}^{J}
\mathbb E_{q(\btheta)}\!\left[
\log p_j\!\left(y_j(s_i)\mid \boldsymbol{\eta}^{(j)}_L(\bs_i;\btheta)\right)
\right].
\end{equation}

The expectation in \eqref{eq:elbo_expand_new} is intractable.
A Monte Carlo estimator is obtained by drawing $\btheta^{(m)}\sim q(\btheta)$, $m=1,\ldots,M$,
which corresponds to sampling dropout masks in each layer and applying them to the variational mean parameters.
Let $\eta_j^{(m)}(s_i)=\eta_j(s_i;\btheta^{(m)})$ denote the $m$th stochastic forward pass.
Then
\begin{equation}\label{eq:elbo_mc_new}
\mathcal L_{\mathrm{MC}}
=
\frac{1}{M}\sum_{m=1}^{M}\sum_{i=1}^{N}\sum_{j=1}^{J}
\log p_j\!\left(y_j(s_i)\mid \boldsymbol{\eta}^{(m,j)}_L(\bs_i)\right)
-
\mathrm{KL}\!\left(q(\btheta)\,\|\,p(\btheta)\right).
\end{equation}

The KL term in \eqref{eq:elbo_mc_new} admits a tractable approximation under the dropout-induced variational family. The KL divergence decomposes additively across layers and units,
\begin{equation*}\label{eq:kl_decomp}
\mathrm{KL}\!\left(q(\btheta)\,\|\,p(\btheta)\right)
=
\sum_{\ell=1}^{L-1}\sum_{a=1}^{k_\ell}
\mathrm{KL}\!\left(q(\bw_{\ell,a},b_{\ell,a})\,\|\,p(\bw_{\ell,a},b_{\ell,a})\right)
+
\sum_{j=1}^{J}\sum_{a=1}^{k_L^{(j)}}
\mathrm{KL}\!\left(q(\bw_{L,a}^{(j)},b_{L,a}^{(j)})\,\|\,p(\bw_{L,a}^{(j)},b_{L,a}^{(j)})\right).
\end{equation*}
Following the standard MC-dropout variational interpretation, the mixture-normal KL is approximated by a quadratic penalty on the variational means.
Specifically, up to an additive constant that does not depend on $\{\bmu^w_{\ell,a},\mu^b_{\ell,a}\}$,
\begin{align}
\mathrm{KL}\!\left(q(\btheta)\,\|\,p(\btheta)\right)
&\approx
\frac{1}{2}\sum_{\ell=1}^{L-1} p_\ell
\left(
\|\bmu^w_{\ell}\|_F^2+\|\bmu^b_{\ell}\|_2^2
\right)
+
\frac{1}{2}\sum_{j=1}^{J} p_L^{(j)}
\left(
\|\bmu^{w,(j)}_{L}\|_F^2+\|\bmu^{b,(j)}_{L}\|_2^2
\right),
\label{eq:kl_weightdecay}
\end{align}
where $\bmu^w_{\ell}\in\mathbb R^{k_\ell\times k_{\ell-1}}$ and $\bmu^b_{\ell}\in\mathbb R^{k_\ell}$ collect the shared-layer variational means, and $\bmu^{w,(j)}_{L}$ and $\bmu^{b,(j)}_{L}$ collect the corresponding variational means for the $j$th output head. The approximation \eqref{eq:kl_weightdecay} implies that the negative ELBO is equivalent to the deterministic training objective in \eqref{eq:nn-loss} when the layer-wise weight decay parameters are chosen to match the KL penalty, namely $\lambda_\ell = p_\ell/2N$ and $\lambda_L^{(j)} = p_L^{(j)}/2N$. Consequently, optimizing the deterministic neural network with dropout can be viewed as maximizing an ELBO under a dropout-induced variational family. 

Under MC dropout, resampling Bernoulli masks across layers induces stochastic parameter draws from the dropout variational approximation~\citep{gal2016bayesian}. In particular, a draw can be represented as $\btheta^{(m)}=\hat{\btheta}\odot\br^{(m)}$ (equivalently, by applying the corresponding node masks to the weight rows and biases), where $\br^{(m)}$ collects the dropout masks across all layers and heads. Given $\btheta^{(m)}$, a stochastic forward pass through the network yields the natural-parameter realization $\eta_L^{(m,j)}(s)=\eta_L(s;\btheta^{(m,j)})$, following the masked forward propagation in Algorithm~\ref{alg:multideepgp_train_pred}.

% With $\widehat{\btheta}$ denoting the optimized variational means obtained from minimizing \eqref{eq:nn-loss}, approximate posterior samples are generated by repeating the following dropout resampling procedure for $m=1,\ldots,M$. For each shared layer $\ell=1,\ldots,L-1$, draw a node mask
% \[
% \bz_\ell^{(m)}=(z_{\ell,1}^{(m)},\ldots,z_{\ell,k_\ell}^{(m)})^\top,
% \qquad
% z_{\ell,a}^{(m)}\sim\mathrm{Bernoulli}(p_\ell),
% \]
% where $p_\ell\in[0,1]$ is the keep probability (equivalently, dropout rate $1-p_\ell$). A stochastic parameter draw for the shared layer is then defined by
% \begin{equation*}\label{eq:drop_shared}
% \bW_\ell^{(m)}=\mathrm{Diag}(\bz_\ell^{(m)})\,\hat{\bW}_\ell,
% \qquad
% \bb_\ell^{(m)}=\bz_\ell^{(m)}\odot \hat{\bb}_\ell.
% \end{equation*}
% For each outcome-specific output head $j=1,\ldots,J$, draw
% \[
% \bz_{L}^{(m,j)}=(z_{L,1}^{(m,j)},\ldots,z_{L,k_L^{(j)}}^{(m,j)})^\top,
% \qquad
% z_{L,a}^{(m,j)}\sim\mathrm{Bernoulli}(p_L^{(j)}),
% \]
% with $p_L^{(j)}\in[0,1]$, and define
% \begin{equation*}\label{eq:drop_head}
% \bW_{L}^{(m,j)}=\mathrm{Diag}(\bz_{L}^{(m,j)})\,\hat{\bW}_{L}^{(j)},
% \qquad
% \bb_{L}^{(m,j)}=\bz_{L}^{(m,j)}\odot \hat{\bb}_{L}^{(j)}.
% \end{equation*}
% All such masked weights and biases across the shared layers and the $J$ output heads together define a stochastic parameter draw $\btheta^{(m)}$ from the dropout variational approximation; the resulting training and prediction procedure is summarized in Algorithm~\ref{alg:multideepgp_train_pred}.

\begin{algorithm}[H]
\scriptsize 
\caption{MultiDeepGP}
\label{alg:multideepgp_train_pred}
\DontPrintSemicolon
\setlength{\algomargin}{1em}   
\setlength{\interspacetitleruled}{1pt}
\setlength{\algotitleheightrule}{0.6pt}
\setlength{\algoheightrule}{0.6pt}

\KwIn{
Spatial locations $\{\bs_i\}_{i=1}^N$;
mixed outcomes $\{\by(\bs_i)\}_{i=1}^N$ with $\by(\bs_i)=(y_1(\bs_i),\ldots,y_J(\bs_i))^\top$;
dropout probabilities $\{p_\ell\}_{\ell=1}^{L-1}$; learning rate $\rho$; number of MC draws $M$.
}
\KwOut{
Trained parameters $\widehat{\btheta}$; for a new location $\bs^\ast$: predictive distribution.
}

\vspace{2pt}
\tcp{\textbf{Part I. Training MultiDeepGP}}
\BlankLine
\textbf{Step 1: Initialize parameters.}\;
Initialize $\btheta^{(0)}=\Big\{\{(\bW_\ell,\bb_\ell)\}_{\ell=1}^{L-1},\{(\bW_L^{(j)},b_L^{(j)})\}_{j=1}^J\Big\}$\;

\BlankLine
\textbf{Stochastic forward pass with dropout.}\;
\For{$i=1,\ldots,N$}{
Set $\bphi_{i,0}=\bs_i$\;
\For{$\ell=1,\ldots,L-1$}{
Sample dropout mask $\br_\ell \sim \text{Bernoulli}(1-p_\ell)$\;
Compute shared hidden feature
$\bphi_{i,\ell}=\sigma_\ell(\bW_\ell\bphi_{i,\ell-1}+\bb_\ell)\odot\br_\ell$.
}
Define the shared representation $H(\bs_i)=\bphi_{i,L-1}$\;

\textbf{Outcome-specific heads and mixed likelihood.}\;
\For{$j=1,\ldots,J$}{
Compute linear predictor and map to the natural parameter $\eta^{(j)}_L(\bs_i)=\sigma_L^{(j)}\!\big(\bW_L^{(j)} H(\bs_i) + b_L^{(j)}\big)$
}
}

\BlankLine
\textbf{Objective and parameter update.}\;
Minimize the regularized negative log-likelihood in Eq.~\eqref{eq:nn-loss} and update $\btheta \leftarrow \btheta - \rho \nabla_\btheta \mathcal{L}(\btheta)$\;

\BlankLine
Set $\widehat{\btheta}\leftarrow \btheta$\;

\BlankLine
\tcp{\textbf{Part II. Prediction and uncertainty quantification (MC dropout)}}
\BlankLine
\KwIn{New location $\bs^\ast$ and $\widehat{\btheta}$}\;
\For{$m=1,\ldots,M$}{
Set $\bphi_{0}^{(m)}(\bs^\ast)=\bs^\ast$\;

\For{$\ell=1,\ldots,L-1$}{
Sample node mask $\bz_\ell^{(m)} \sim \text{Bernoulli}(1-p_\ell)$\;
Define masked parameters
$
\bW_\ell^{(m)}=\mathrm{Diag}(\bz_\ell^{(m)})\,\widehat{\bW}_\ell,
\qquad
\bb_\ell^{(m)}=\bz_\ell^{(m)}\odot \widehat{\bb}_\ell
$

Compute shared hidden feature $\bphi_{\ell}^{(m)}(\bs^\ast)
=
\sigma_\ell\!\big(\bW_\ell^{(m)} \bphi_{\ell-1}^{(m)}(\bs^\ast)+\bb_\ell^{(m)}\big)$
}
Define $H^{(m)}(\bs^\ast)=\bphi_{L-1}^{(m)}(\bs^\ast)$\;

\For{$j=1,\ldots,J$}{
Sample output-head mask $\bz_{L}^{(m,j)} \sim \text{Bernoulli}(1-p_L^{(j)})$\;
Define masked head parameters $\bW_{L}^{(m,j)}=\mathrm{Diag}(\bz_{L}^{(m,j)})\,\widehat{\bW}_{L}^{(j)},
\qquad
b_{L}^{(m,j)}=\bz_{L}^{(m,j)}\odot \widehat{b}_{L}^{(j)}$

Compute natural parameter $\eta_{j}^{(m)}(\bs^\ast)
=
\sigma_{L}^{(j)}\!\big(\bW_{L}^{(m,j)} H^{(m)}(\bs^\ast)+ b_{L}^{(m,j)}\big)$

}
}

\textbf{Predictive distribution.}\;
\[
p\!\left(\by(\bs^\ast)\mid \mathcal D\right)
\approx
\frac{1}{M}\sum_{m=1}^M
\prod_{j=1}^J
p\!\left(y_j(\bs^\ast)\mid \eta_L^{(m,j)}(\bs^\ast)\right).
\]
\end{algorithm}

\subsection{Predictive distribution}

For a new location $\bs^\ast$, the posterior predictive distribution is
obtained by marginalizing over the network parameters,
\begin{equation}
p\!\left(\by(\bs^\ast)\mid \mathcal D\right)
=
\int
\prod_{j=1}^J
p\!\left(y_j(\bs^\ast)\mid \eta_j(\bs^\ast;\btheta)\right)
\,\pi(\btheta\mid\mathcal D)\,d\btheta,
\label{eq:ppd}
\end{equation}
where $\eta_j(\bs^\ast;\btheta)$ is the outcome-specific linear predictor that
indexes the conditional distribution of $y_j(\bs^\ast)$ through the link
function $\phi_j(\cdot)$. Since the posterior $\pi(\btheta\mid\mathcal D)$ is intractable, we approximate
Eq.~\eqref{eq:ppd} using Monte Carlo dropout~\citep{gal2016dropout}.
Let $\btheta^{(t)}\sim q(\btheta)$ denote the effective network parameters
induced by the $t$th stochastic forward pass. The predictive distribution is then approximated by
\begin{equation}
p\!\left(\by(\bs^\ast)\mid \mathcal D\right)
\approx
\frac{1}{M}\sum_{m=1}^M
\prod_{j=1}^J
p\!\left(y_j(\bs^\ast)\mid \eta_L^{(m,j)}(\bs^\ast)\right),
\quad
\eta_L^{(m,j)}(\bs^\ast)=\eta_L(\bs^\ast;\btheta^{(m,j)}).
\label{eq:ppd-mc}
\end{equation}

The conditional distribution of each outcome is specified according to its
type, with the likelihood
$p\!\left(y_j(\bs^\ast)\mid \eta_j(\bs^\ast;\btheta)\right)$
chosen accordingly. For binary outcomes, we use a Bernoulli likelihood with logistic link, $y_j(\bs^\ast)\mid \eta_j \sim \mathrm{Bernoulli}\!\left(\mathrm{logit}^{-1}(\eta_j)\right)$. For count outcomes, we adopt a Poisson likelihood with log link, $y_j(\bs^\ast)\mid \eta_j \sim \mathrm{Poisson}\!\left(\exp(\eta_j)\right).$ For continuous outcomes, we assume a Gaussian likelihood with identity link, $y_j(\bs^\ast)\mid \eta_j \sim \mathcal N\!\left(\eta_j,\sigma_j^2\right)$ where  $\sigma_j^2$ is estimated empirically from the training data using the fitted residuals, $\hat\sigma_j^2
= \frac{1}{N_j}\sum_{i:\, y_{ip}\ \mathrm{obs}}
(y_{ij}-\hat\eta_{ip})^2,
\hat\eta_{ip}=\eta_j(\bs_i;\hat\btheta),$ where $N_j$ denotes the number of observed training samples for outcome $p$.

\section{Simulation Study}
In this section, we validate the proposed model through two simulation designs. The first considers a one dimensional stationary Gaussian process with mixed outcome types, while the second examines a two dimensional nonstationary surface with mixed outcomes. Together, these simulation settings allow us to assess the model to capture both nonstationary spatial structure and dependence across mixed outcome types through shared latent representations.

\subsection{Competitors and Metric of Comparison}

To assess the predictive performance of the proposed model, we compare it with two competing approaches that represent commonly used alternatives for modeling mixed spatial outcome data. Our proposed method, denoted as MultiDeepGP, serves as the primary benchmark and explicitly models dependence across mixed outcomes through a shared latent spatial structure. The first competing approach, referred to as MultiDNN, is a deterministic deep neural network baseline. In this approach, spatial locations are included as input covariates through their geographic coordinates, but no explicit spatial process or latent spatial component is modeled. This baseline is included to isolate the contribution of the proposed framework in incorporating shared latent spatial structure and likelihood-based uncertainty quantification, beyond what can be achieved by purely deterministic deep learning models that use spatial coordinates as inputs. The second competitor is spatial kriging denoted as Kriging, a classical geostatistical method designed for spatially indexed data \citep{cressie1993statistics, banerjee2014hierarchical}. Because kriging is defined for single response processes and does not naturally extend to mixed outcome types, we fit separate kriging models for each outcome. Specifically, for the binary outcome, we apply indicator kriging, while for the count and continuous outcomes, standard Gaussian kriging is used after appropriate transformations. This comparison evaluates the extent to which a traditional spatial smoothing approach can capture spatial structure in the absence of joint modeling across outcome types.

Performance is evaluated using metrics tailored to each outcome type. For the binary outcome, predictive accuracy is assessed using the area under the receiver operating characteristic curve (AUC) and the Brier score; AUC quantifies discrimination by measuring the model’s ability to distinguish between classes independent of any classification threshold, while the Brier score measures the mean squared difference between predicted probabilities and observed outcomes, providing a proper scoring rule that jointly reflects calibration and overall probabilistic accuracy. For the count and continuous outcomes, we report the root mean squared error (RMSE) to quantify point prediction accuracy. In addition, for models that provide predictive distributions, we evaluate uncertainty quantification using empirical coverage probabilities of nominal 95\% prediction intervals and the corresponding average interval widths. These metrics allow us to assess not only predictive accuracy but also the calibration and sharpness of predictive uncertainty. We implement our MultiDeepGP approach using {\tt{TensorFlow}}, an open-source machine learning platform. All experiments were conducted on a 4-core, 2.3 GHz Intel Core i7 processor. 

\subsection{Simulation Case 1: One Dimensional Stationary Gaussian Process with Mixed Outcomes}

We consider a one dimensional spatial domain $s \in [0,1]$ and generate data from a stationary Gaussian process to evaluate model performance under a controlled spatial dependence structure. \textcolor{red}{The simulation setup closely follows the setting in \citet{chen2024deepkriging} as a representative benchmark for deep learning–based spatial models, with an extension to accommodate both continuous and binary outcomes.}

Let $z(s)$ denote a latent spatial process defined as
\[z(s) = \mu + \nu(s) + \epsilon(s),\]
where $\mu = 1$ is a constant mean term, $\nu(s)$ is a zero mean Gaussian process, and $\epsilon(s)$ represents independent measurement noise. The spatial process $\nu(s)$ is modeled with an exponential covariance function
\[\text{Cov}\{\nu(s), \nu(s')\}
= \sigma^2 \exp\left( -\frac{|s - s'|}{\rho} \right),\]
with variance $\sigma^2 = 1$ and range parameter $\rho = 0.1$. The nugget effect is modeled as $\epsilon(s) \sim \mathcal{N}(0, \tau^2)$ with $\tau^2 = 0.01$, independent across spatial locations. We generate $N = 1000$ spatial locations equally spaced over the unit interval $[0,1]$. For each simulated dataset, 800 locations are randomly selected as training data, and the remaining 200 locations are reserved for testing. This procedure is repeated independently for 100 simulation replicates.

The continuous response is generated directly from the latent spatial process as $Y_{\text{cont}}(s) = z(s)$, corresponding to a noise contaminated spatial signal with smooth dependence induced by the underlying Gaussian process. To generate a binary response that shares the same latent spatial structure, we map the latent process $z(s)$ to a Bernoulli probability through a logistic transformation, $p(s) = \text{logit}^{-1}\left( \frac{z(s) - c}{\kappa} \right)$ where $c$ is a centering constant and $\kappa > 0$ controls the steepness of the logistic function. In our experiments, we set $c = \mu=1$ and $\kappa = 0.35$ to obtain approximately balanced binary classes. The binary response is then generated as $Y_{\text{bin}}(s) \sim \text{Bernoulli}\{p(s)\}$. To incorporate a count response that also depends on the same latent spatial process, we specify a Poisson model with a log link, $Y_{\text{count}}(s) \sim \text{Poisson}\{\lambda(s)\}, \quad 
\log \lambda(s) = \alpha + \beta\, z(s),$ where $\alpha$ controls the overall event rate and $\beta$ governs the strength of association between the latent spatial field and the count intensity. Here we set $\alpha=-0.25$ and $\beta=0.60$. This construction ensures that all three outcomes share a common latent spatial structure while exhibiting distinct data generating mechanisms appropriate to their respective outcome types.

% \textcolor{red}{Classical co-kriging is fundamentally restricted to continuous responses due to its reliance on linear minimum-variance predictors and second-order moment structures. Binary outcomes violate these assumptions, as their likelihood is inherently non-Gaussian and cannot be represented through linear predictors with well-defined stationary cross-covariance functions. While indicator kriging has been used as a heuristic baseline, it lacks a coherent probabilistic interpretation and does not provide joint uncertainty quantification. Fully principled joint modeling of continuous and binary spatial outcomes requires latent Gaussian process models with mixed likelihoods, which necessitate approximate Bayesian inference and substantially increase computational complexity. As a result, such models are largely absent from the classical kriging literature and have only recently been explored within the broader Bayesian and machine learning communities.}

\begin{table}[h!]
\centering
\caption{This table reports predictive performance for binary, count, and continuous outcomes under Simulation Case 1, corresponding to a one-dimensional stationary Gaussian process. Results are averaged over 100 independent simulation replicates, with standard deviations shown in parentheses. Computational time is reported in seconds. For each metric, the best performing method is highlighted in bold.}
\label{tab:performance1}
\begin{tabular}{@{}llccc@{}}
\hline
Case & Metric & MultiDeepGP & MultiDNN & Kriging \\
\hline
Binary Case 
& AUC 
& 0.871 (0.04) 
& 0.723 (0.11) 
& \textbf{0.879} (0.04) \\
& Brier Score 
& 0.134  (0.02) 
& 0.184  (0.04)  
& \textbf{0.130}  (0.03) \\ 
\hline
\multirow{3}{*}{Count Case}
& RMSE 
& 1.378 (0.34) 
& 1.487 (0.28)
& \textbf{1.341} (0.18) \\
& Coverage 
& 0.935 (0.08) 
& -- 
& \textbf{0.962} (0.02) \\
& Interval 
&  \textbf{3.910} (0.67) 
& -- 
& 4.407 (0.77) \\
\hline
\multirow{3}{*}{Continuous Case}
& RMSE 
& 0.307 (0.03) 
& 0.693 (0.14) 
& \textbf{0.172} (0.02) \\
& Coverage 
& \textbf{0.958} (0.01) 
& -- 
& 0.910 (0.09) \\
& Interval 
& 1.339 (0.11) 
& -- 
& \textbf{0.794} (0.45) \\
\hline
Time (secs) 
&  
& 19.49 (2.29) 
& 15.29 (1.52) 
& 21.93 (9.29) \\
\hline
\end{tabular}
\end{table}

Table~\ref{tab:performance1} summarizes predictive accuracy and uncertainty quantification across binary, count, and continuous outcomes. For the binary case, MultiDeepGP achieves an AUC of 0.871, substantially outperforming the MultiDNN baseline and closely matching the performance of Kriging used as an indicator-based benchmark. In the count setting, MultiDeepGP attains lower RMSE than MultiDNN and predictive accuracy comparable to Kriging, while providing well-calibrated uncertainty with empirical 95\% coverage near the nominal level and narrower predictive intervals. In contrast, MultiDNN does not provide uncertainty estimates and is therefore omitted from coverage and interval comparisons.

For continuous outcomes, Kriging yields the lowest RMSE, consistent with its theoretical optimality under correctly specified Gaussian process covariance functions~\citep{chen2024deepkriging}. Nevertheless, MultiDeepGP maintains competitive point prediction performance and achieves coverage closer to the nominal level than kriging, while providing reliable uncertainty quantification across all outcome types. It is worth noting that the strong performance of kriging in this setting is consistent with the approximately stationary spatial structure of the underlying data, under which classical kriging methods are known to perform well. In contrast, MultiDeepGP does not rely on stationarity assumptions and remains effective in more general settings. Importantly, MultiDeepGP offers a favorable trade-off between well-calibrated uncertainty quantification and computational efficiency, with runtime comparable to MultiDNN and substantially faster than kriging, highlighting its practicality for joint modeling of heterogeneous spatial responses.

% \textcolor{red}{When the underlying data-generating process is a Gaussian process with a correctly specified covariance function, Kriging is theoretically optimal in the sense of minimizing the mean squared prediction error. Consistent with this theoretical property, as shown in \citet{chen2024deepkriging}, the simulation study in one-dimensional Gaussian process settings demonstrates that Kriging with the true covariance function achieves the best predictive performance among all competing methods. However, despite being slightly less optimal in terms of point prediction accuracy, our model achieves adequate and well-calibrated 95\% predictive coverage compared to Kriging, reflecting its ability to effectively quantify predictive uncertainty even when the covariance structure is not explicitly specified.}

\subsection{Simulation Case 2: Two Dimensional Nonstationary Surface with Mixed Outcomes}

We next consider a two dimensional nonstationary simulation setting follows the setting in \citet{chen2024deepkriging}, with an extension to accommodate both count and binary outcomes. The goal of this experiment is to assess the ability of competing models to capture spatially varying smoothness and complex nonlinear patterns in higher dimensional spatial domains.

Let $\bs = (s_x, s_y)^{\top} \in [0,1]^2$ denote a two dimensional spatial location and define $\bar{s} = (s_x + s_y)/2$. The true underlying process is given by
\[f(\bs) =
\sin\!\left\{30(\bar{s} - 0.9)^4\right\}
\cos\!\left\{2(\bar{s} - 0.9)\right\}
+ \frac{\bar{s} - 0.9}{2}.\]
This function behaves differently across space. When $\bar{s}$ is far from 0.9, it fluctuates quickly with noticeable local peaks and valleys, while when $\bar{s}$  is close to 0.9, it varies more slowly and appears much smoother. We sample $N = 900$ spatial locations within the $[0,1]^2$. For each simulated dataset, $80\%$ of the locations are randomly selected as training data, and the remaining $20\%$ are reserved for testing. This sampling procedure is repeated independently for 100 simulation replicates.

To generate responses of different types that share the same underlying nonstationary spatial structure, we define a unified linear predictor $\eta(\bs)=\alpha+\beta f(\bs)$ where $\alpha$ controls the overall intensity or baseline level of the response and $\beta$ governs the spatial contrast induced by $f(\bs)$. Different response types are then obtained by applying appropriate link functions to this shared linear predictor. For count data, we employ a Poisson model with a log link. Specifically, the Poisson mean is defined as $\lambda(\bs) = \exp(\eta(\bs))$, and the observed response is generated as $Y_{\text{pois}} \sim \text{Poisson}(\lambda(\bs))$. For binary data, we use a Bernoulli model with a logistic link. The success probability is given by $p(\bs)=\text{logit}^{-1}(\eta(\bs))$, and the binary outcome is generated as $Y_{\text{bin}} \sim \text{Bernoulli}(p(\bs))$. For continuous data, we directly link the response to the latent spatial signal through an additive noise model, $Y_{\text{cont}}(\bs) = \eta(\bs) + \epsilon$ where $\epsilon \sim N(0,\sigma^2)$. Throughout all experiments, we fix the hyperparameters to $\alpha=0.5, ~ \beta=3$ and $\sigma^2 = 0.25$.

Table~\ref{tab:performance2} summarizes predictive performance for Simulation Case 2, corresponding to a two-dimensional nonstationary Gaussian process with mixed outcome types. Overall, MultiDeepGP consistently outperforms competing methods across binary, count, and continuous outcomes, achieving strong predictive accuracy together with well-calibrated uncertainty. These results highlight the advantage of jointly modeling mixed outcomes through a shared latent spatial representation, allowing the model to flexibly adapt to complex spatial heterogeneity. Compared to MultiDeepGP, MultiDNN exhibits weaker performance across outcome types, particularly in the count and continuous cases. While the deep neural network is capable of capturing nonlinear relationships through its nested layer structure, it treats each outcome largely independently and does not explicitly enforce a shared latent spatial structure. This limits its ability to coherently exploit cross-outcome dependence induced by a common underlying spatial process. Moreover, as a deterministic modeling approach, MultiDNN does not provide principled uncertainty quantification, and therefore cannot deliver predictive coverage or interval estimates, further constraining its utility in spatial settings where uncertainty assessment is essential.

In contrast to Simulation Case 1, where Kriging performs competitively under a one-dimensional stationary setting, the performance of Kriging deteriorates markedly in this two-dimensional nonstationary regime. When the underlying spatial process exhibits nonstationary dependence, stationary covariance-based kriging can lead to degraded predictive accuracy and miscalibrated uncertainty \citep{sampson1992nonparametric}. These issues are further amplified in higher-dimensional settings, where local changes in smoothness and dependence structure are more pronounced, as well as in multivariate and mixed-outcome settings, where specifying valid and flexible nonstationary cross-covariance structures remains a fundamental challenge \citep{gneiting2010matern}. As a result, Kriging either yields overly conservative predictive intervals or suffers from loss of calibration under covariance misspecification.

Taken together, these results underscore the importance of shared latent structure modeling for mixed-outcome spatial data, particularly under nonstationary and higher-dimensional regimes. By combining nonlinear representation learning with a unified probabilistic framework, MultiDeepGP achieves a favorable balance between predictive accuracy, uncertainty calibration, and computational efficiency in settings where classical kriging-based and deterministic deep learning approaches face fundamental limitations.

\begin{table}[h!]
\centering
\caption{This table reports predictive performance for binary, count, and continuous outcomes under Simulation Case 2, corresponding to a two-dimensional nonstationary Gaussian process. Results are averaged over 100 independent simulation replicates, with standard deviations shown in parentheses. Computational time is reported in seconds. For each metric, the best performing method is highlighted in bold.}
\label{tab:performance2}
\begin{tabular}{@{}llccc@{}}
\hline
Case & Metric & MultiDeepGP & MultiDNN & Kriging \\
\hline
Binary Case 
& AUC 
& \textbf{0.705} (0.04) 
& 0.701 (0.05) 
& 0.634 (0.04) \\
& Brier Score 
&  \textbf{0.209} (0.01) 
&  0.212 (0.01)
&  0.242 (0.02) \\
\hline
\multirow{3}{*}{Count Case}
& RMSE 
& \textbf{1.445} (0.11) 
& 1.504 (0.19) 
& 1.846 (0.11) \\
& Coverage 
& \textbf{0.974} (0.01) 
& -- 
& 0.995 (0.01) \\
& Interval 
& \textbf{4.898} (0.17) 
& -- 
& 7.796 (0.23) \\
\hline
\multirow{3}{*}{Continuous Case}
& RMSE 
& \textbf{0.536} (0.03) 
& 0.573 (0.09) 
& 0.566  (0.03)  \\
& Coverage 
& \textbf{0.961} (0.01) 
& -- 
& 0.844 (0.03)  \\
& Interval 
& 2.244 (0.08) 
& -- 
& \textbf{1.612} (0.06)  \\
\hline
Time (secs) 
&  
& 24. 091 (1.48) 
& 21.201 (3.10) 
& 42.104 (2.52)  \\
\hline
\end{tabular}
\end{table}

\section{Coupled Environmental–Health Prediction in the African Great Lakes Region}

We analyze georeferenced survey data from the African Great Lakes region collected through the 2015 Demographic and Health Surveys, comprising spatially indexed cluster-level measurements across nine contiguous countries. At each location, the observed responses capture complementary aspects of environmental and public health conditions, including vegetation presence, malaria burden, and water availability. Vegetation is derived from a remotely sensed land surface index and summarized as an indicator of vegetated land cover, malaria incidence is recorded as a nonnegative count reflecting local disease burden, and water availability is measured on a continuous scale. Although these responses follow different marginal distributions (binary, count, and continuous), they are expected to be governed by shared latent spatial processes driven by common environmental factors and regional heterogeneity, with potentially nonstationary spatial structure across the study region.

Formally, at each spatial location $\bs_i$, we observe a multivariate response
\[
\by(\bs_i) = \big(y_{1}(\bs_i),\, y_{2}(\bs_i),\, y_{3}(\bs_i)\big)^\top,
\]
where $y_{1}(\bs_i)$ denotes a vegetation indicator, $y_{2}(\bs_i)$ represents malaria incidence counts, and $y_{3}(\bs_i)$ corresponds to water availability.
Vegetation is summarized using a remotely sensed index, which we binarize for interpretability by classifying values greater than or equal to $0.2$ as vegetated areas and values below $0.2$ as non vegetated. Malaria incidence is observed as a count outcome, while water availability is measured on a continuous scale. In addition, rainfall is included as an exogenous environmental covariate, as it serves as a key climatic driver influencing vegetation dynamics, malaria transmission, and hydrological conditions~\citep{gbaguidi2025environmental}. \textcolor{red}{More generally, such exogenous covariates can be incorporated in our framework by concatenating them with the shared latent representation at the final layer, allowing covariate effects to enter the outcome-specific link functions.}

To assess out of sample predictive performance, we randomly partition the data into a training set of size $N = 3500$ and a test set of size $N_{\text{test}} = 1241$. Spatial dependence is incorporated through a basis-function representation of location. Specifically, we construct a spatial design matrix $\mathbf{X}$ by overlaying a $25\times 25$ regular lattice of candidate knots on the bounding rectangle of the study area and retain the $239$ knots that lie within the African Great Lakes polygon. For each spatial location $\bs$ and knot $\bu_j$, we use thin–plate spline (TPS) radial bases
\[
X_j(\mathbf{s}) \;=\; \|\mathbf{s}-\mathbf{u}_j\|^{2}\,\log\!\big(\|\mathbf{s}-\mathbf{u}_j\|\big), 
\qquad j=1,\dots,239,
\]
where $\|\cdot\|$ is the Euclidean norm. Evaluating these $239$ bases at each observation yields the input design $\mathbf{X}\in\mathbb{R}^{3500\times 239}$, which serves as the common spatial input representation for all models considered.

\begin{table}[h!]
\centering
\caption{Predictive performance and uncertainty quantification results for the real data analysis. We compare MultiDeepGP, a multi-task deep neural network (MultiDNN), and spatial kriging across three outcomes: vegetation indicator (binary), malaria incidence (count), and water availability (continuous).}
\label{tab:realdata}
\begin{tabular}{@{}llccc@{}}
\hline
Case & Metric & MultiDeepGP & MultiDNN & Kriging \\
\hline
vegetation indicator 
& AUC 
& 0.941  
& 0.845  
& \textbf{0.952}   \\
& Brier Score 
& \textbf{0.096}
& 0.185
& 0.112 \\
\hline
\multirow{3}{*}{malaria incidence}
& RMSE 
& \textbf{40.444} 
& 186.446   
& 63.867 \\
& Coverage 
& \textbf{0.833} 
& -- 
& 0.510  \\
& Interval 
& 87.549
& -- 
& \textbf{49.169}   \\
\hline
\multirow{3}{*}{water availability}
& RMSE 
&\textbf{0.219}  
& 0.649  
& 0.615  \\
& Coverage 
& \textbf{0.969}  
& -- 
& 0.985  \\
& Interval 
& \textbf{1.022}
& -- 
& 4.645   \\
\hline
Time (secs) 
&  
&  80.902  
&  42.104 
&  302.910  \\
\hline
\end{tabular}
\end{table}

Table~\ref{tab:realdata} indicates that the proposed MultiDeepGP achieves a favorable balance between predictive accuracy, calibration, and uncertainty quantification across all three outcomes in the real data analysis. For the vegetation indicator (binary outcome), kriging attains the highest AUC, indicating strong discriminatory ability in ranking locations by vegetation presence. However, MultiDeepGP achieves a substantially lower Brier score, reflecting superior probabilistic calibration. This discrepancy highlights an important distinction between discrimination and calibration: while kriging can effectively separate positive and negative classes, its predicted probabilities tend to be overconfident or poorly calibrated, whereas MultiDeepGP produces probability estimates that more accurately reflect empirical frequencies. This behavior is consistent with kriging’s reliance on Gaussian assumptions and fixed covariance structures, which are not tailored to binary outcomes. For the malaria incidence (count outcome), MultiDeepGP substantially outperforms both MultiDNN and kriging in terms of predictive accuracy, achieving the lowest RMSE. Moreover, MultiDeepGP provides well-calibrated uncertainty quantification, with empirical coverage close to the nominal 95\% level and moderate prediction interval length. In contrast, kriging exhibits severe undercoverage, suggesting that its uncertainty estimates fail to capture the pronounced nonstationarity and overdispersion present in malaria incidence data. The MultiDNN baseline does not provide uncertainty estimates for count or continuous outcomes, underscoring a key limitation of non-Bayesian deep learning approaches in this setting.

For water availability (continuous outcome), MultiDeepGP again achieves the lowest RMSE among the competing methods, while maintaining coverage close to the nominal level with substantially shorter prediction intervals than kriging. Although kriging attains slightly higher coverage, this comes at the cost of overly wide intervals, indicating conservative and inefficient uncertainty quantification. In contrast, MultiDeepGP delivers sharper predictive intervals while preserving reliable coverage, reflecting a more effective representation of spatially varying uncertainty.

Overall, these results demonstrate that MultiDeepGP consistently provides accurate predictions together with well-calibrated and efficient uncertainty quantification. A key contributor to this performance is the shared latent spatial representation across outcomes, which enables MultiDeepGP to borrow strength across related environmental and health processes. In contrast, MultiDNN models each outcome independently without shared structure, limiting its ability to capture cross-outcome dependencies inherent in the data. This advantage is particularly important in the real data context of coupled environmental–health prediction in the African Great Lakes region, where vegetation, water availability, and malaria incidence are strongly interrelated through complex spatial and ecological mechanisms. By jointly modeling these outcomes, MultiDeepGP more effectively captures shared spatial heterogeneity and yields improved predictive and inferential performance. Additionally, MultiDeepGP achieves these gains with substantially lower computational cost than kriging, further supporting its practical utility for large-scale spatial analyses.

\begin{figure}[htbp]
\begin{center}
\includegraphics[width = 0.5\textwidth]{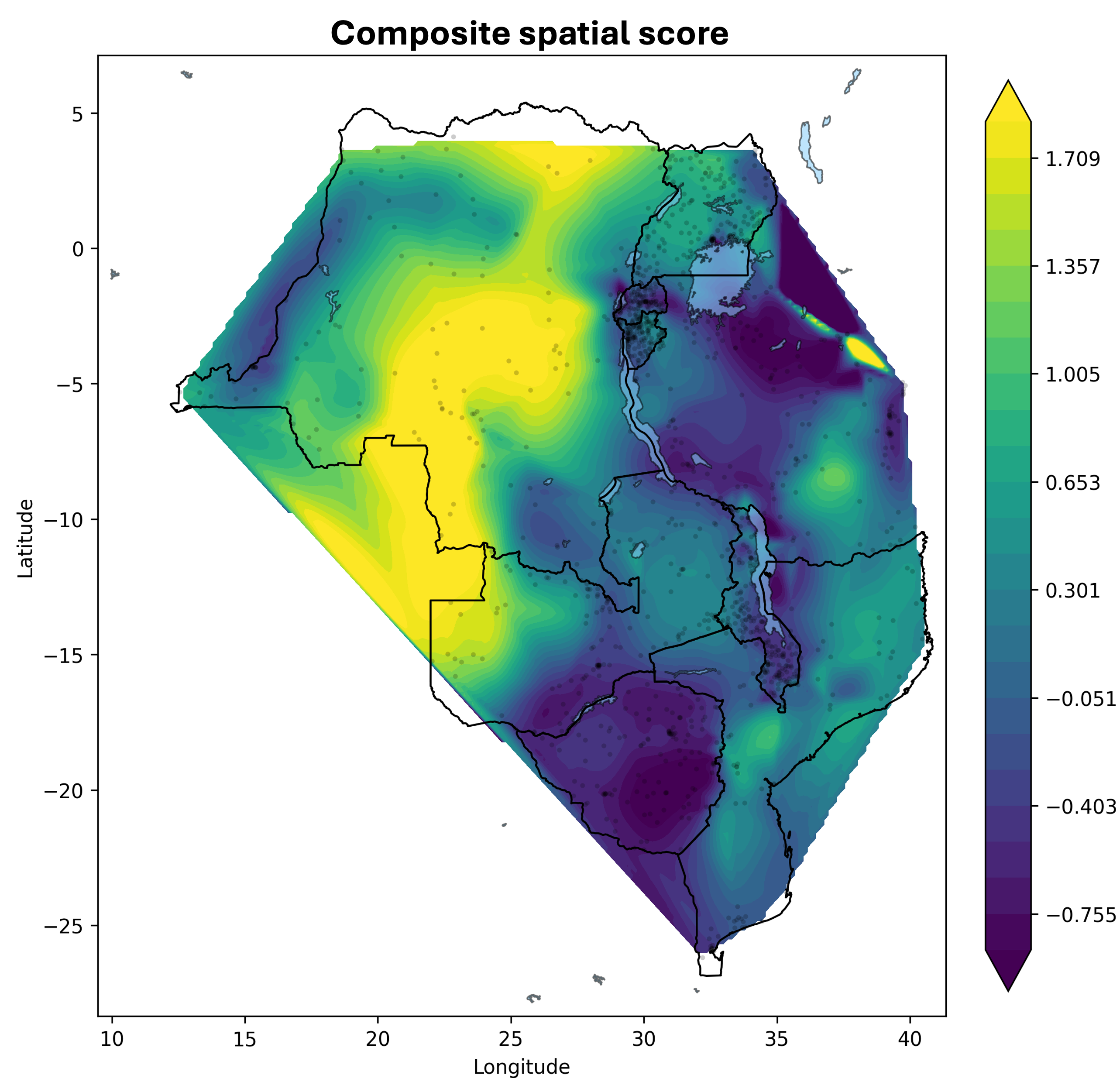}
\end{center}
\caption[]{Shared spatial surface summarizing common latent effects across outcomes.The figure displays a proxy for the shared latent spatial structure learned by MultiDeepGP. Colors indicate the relative strength and direction of the shared spatial influence, with higher values corresponding to locations where all outcomes are jointly elevated relative to their respective means. The surface is obtained by interpolating pointwise predictions at cross-validation locations onto a regular grid for visualization. Black dots denote observation locations.}
\label{Fig:latent}
\end{figure}

To visualize the common spatial pattern shared across outcomes, we construct a composite spatial score by first standardizing the predicted surface of each outcome and then averaging these standardized surfaces across outcomes, following the general principles used in constructing standardized composite indicators~\citep{joint2008handbook}. The resulting surface highlights locations where the outcomes exhibit concordant behavior relative to their respective marginal distributions. In Fig.~\ref{Fig:latent}, positive values indicate regions where vegetation, malaria incidence, and water availability are jointly predicted to be higher than average, whereas negative values correspond to areas where all three outcomes are jointly lower than average.

Regions with high shared spatial scores are concentrated across the western branch of the East African Rift System, particularly along the Albertine Rift and the transitional highland zones bordering the Congo Basin and the Lake Victoria basin. These areas experience elevated seasonal rainfall associated with the migration of the Intertropical Convergence Zone \citep{nicholson2017climate}, and exhibit moderate temperatures due to plateau elevation. Such coupled climatic and topographic conditions promote dense vegetation cover and sustained soil moisture, while also creating environmental suitability for malaria transmission \citep{lindsay1996climate, craig1999climate}. Consequently, vegetation, malaria incidence, and water-related indicators are simultaneously elevated in these regions. These patterns reflect the common spatial structure learned through the shared latent component of MultiDeepGP. This pattern provides visual evidence that joint modeling is appropriate for these data and supports the improved predictive and uncertainty quantification performance observed for MultiDeepGP in Table~\ref{tab:realdata}.

% These patterns reflect the common spatial structure learned through the shared latent component of MultiDeepGP. By borrowing strength across outcomes, the model identifies locations where multiple ecological and health processes co-vary, rather than capturing outcome-specific fluctuations. The resulting surface therefore summarizes cross-outcome spatial coherence, providing a compact representation of their joint spatial behavior. Areas with comparatively low shared scores are generally located in relatively drier or more elevated inland zones, where vegetation productivity, surface water persistence, and malaria suitability are jointly reduced. 

\begin{figure}[htbp]
\centering
\begin{subfigure}[t]{0.95\textwidth}
\centering
\includegraphics[width=\textwidth]{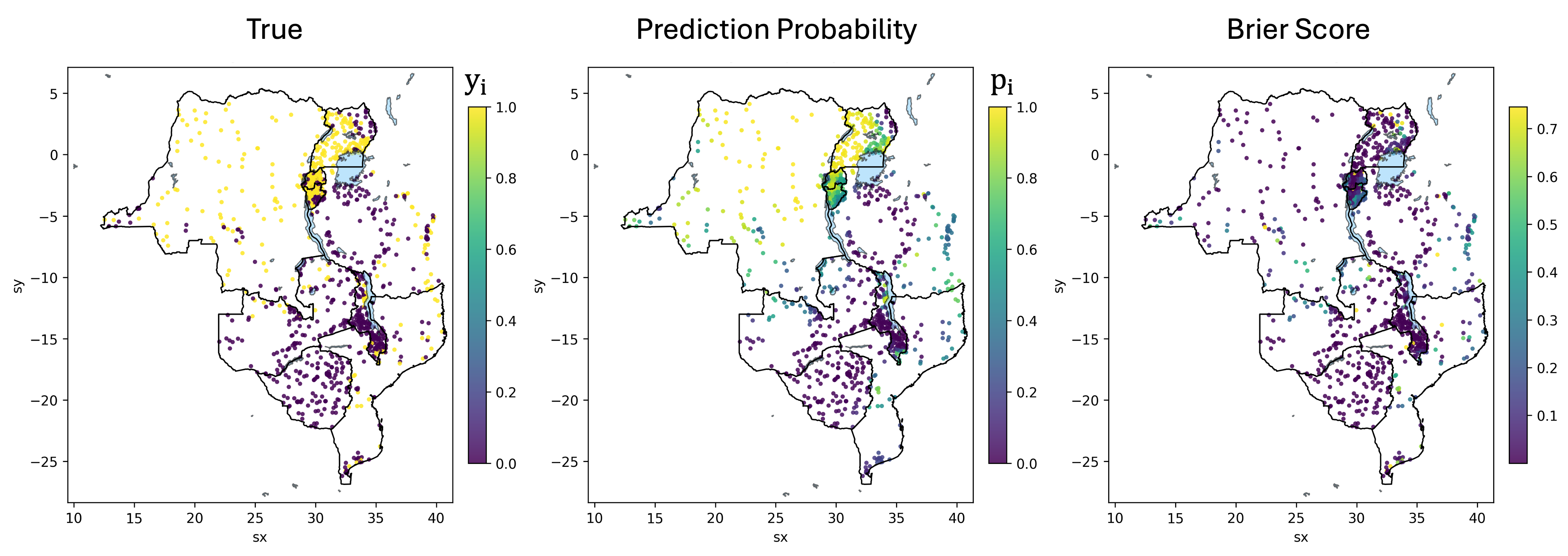}
\caption{ Vegetation indicator (Binary outcome). Pointwise maps at cross-validation locations showing the observed label, the MultiDeepGP predictive probability, and the corresponding pointwise Brier loss $(\by(\bs)-\widehat \bp(\bs))^2$.}
\label{Fig:real_binary}
\end{subfigure}
\vspace{0.4cm}
\begin{subfigure}[t]{0.95\textwidth}
\centering
\includegraphics[width=\textwidth]{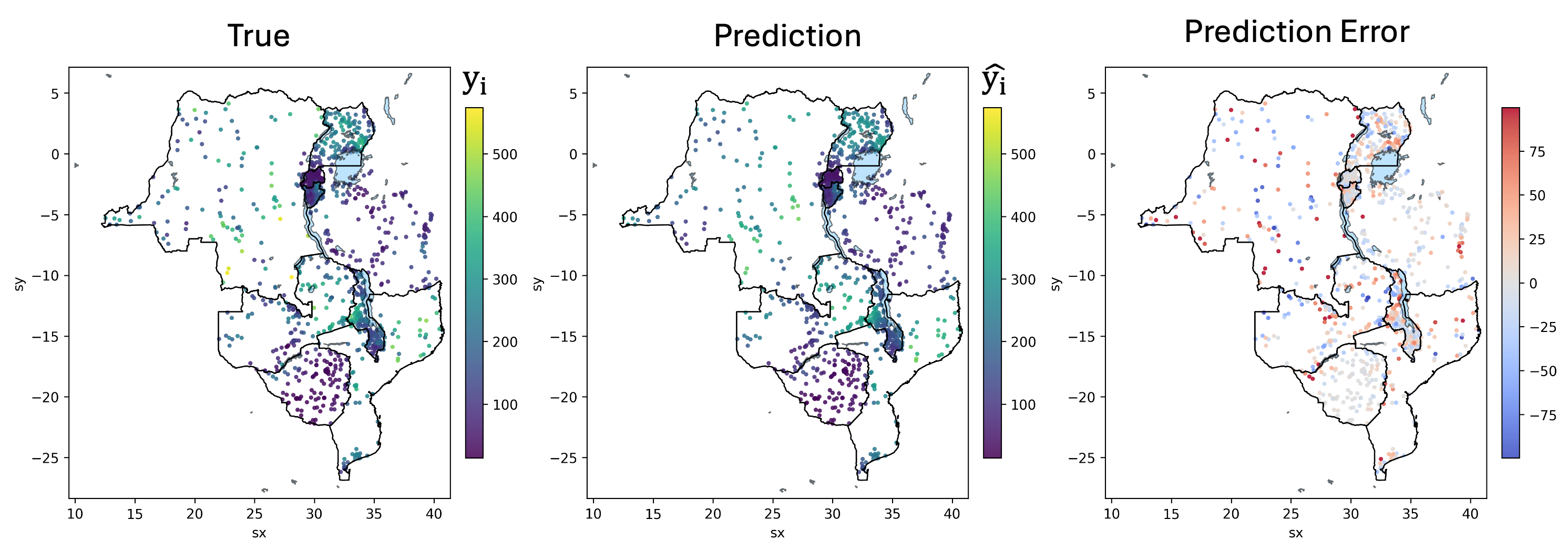}
\caption{Malaria incidence (Count outcome). Pointwise maps at cross-validation locations showing the observed count, the MultiDeepGP predictive mean intensity $\widehat\by(\bs)$,  and the prediction error $\by-\widehat\by(\bs)$.}
\label{Fig:real_count}
\end{subfigure}
\vspace{0.4cm}
\begin{subfigure}[t]{0.95\textwidth}
\centering
\includegraphics[width=\textwidth]{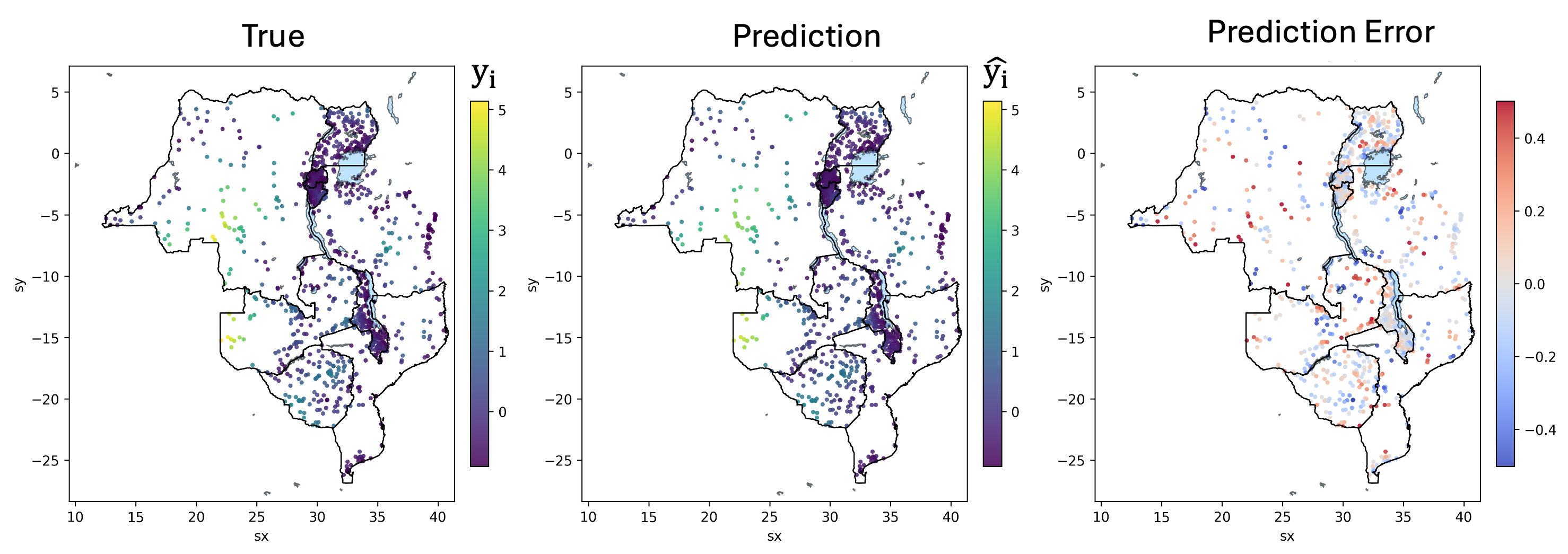}
\caption{ Water availability (continuous outcome). Pointwise maps at cross-validation locations showing the observed response, the MultiDeepGP predictive mean, and the prediction error $\by(\bs)-\widehat \by(\bs)$.}
\label{Fig:real_cont}
\end{subfigure}
\caption{Pointwise predictive diagnostics for the African Great Lakes real data.
Each panel displays MultiDeepGP predictions evaluated at cross-validation locations.
From top to bottom: binary vegetation indicator, malaria incidence (count), and water availability (continuous).
The plots enable a direct spatial comparison between observed outcomes, predictive means, and local prediction errors across heterogeneous outcome types.}
\label{Fig:real_pointwise_diagnostics_vertical}
\end{figure}

Fig.~\ref{Fig:real_pointwise_diagnostics_vertical} presents pointwise predictive diagnostics for the African Great Lakes data across vegetation indicator (binary), malaria incidence (count), and water availability (continuous). For all three tasks, the predictive means closely align with the observed responses at cross-validation locations, indicating that MultiDeepGP effectively captures the dominant spatial structure of each outcome. The corresponding error maps exhibit no discernible large-scale spatial clustering, suggesting the absence of systematic regional bias. This pattern is particularly notable for the count and continuous outcomes, where residuals remain spatially diffuse despite substantial heterogeneity in the observed signals. Together, these diagnostics support good local calibration and stable spatial representation across heterogeneous likelihoods.

Beyond predictive alignment, the diagnostics reveal scientifically meaningful spatial patterns across outcomes. For the vegetation indicator, predictive probabilities concentrate in humid, lake-adjacent zones surrounding Lake Victoria and Lake Tanganyika, particularly across western Kenya, northern Tanzania, and southern Uganda, reflecting well-documented ecological gradients in the East African Rift system~\citep{nicholson2017climate}. For malaria incidence, elevated intensities align with low-lying and water-proximal regions in the Lake Victoria basin and parts of western Uganda, consistent with established links between hydrology, surface water persistence, and mosquito breeding habitats~\citep{hay2005urbanization}. Importantly, prediction errors remain spatially scattered rather than regionally clustered, suggesting that the model captures dominant environmental drivers without introducing systematic regional bias. For water availability, smoother spatial transitions are observed across the Congo Basin and the southern highlands of Tanzania, consistent with large-scale hydrological continuity and basin-level flow structures documented in regional climatology studies~\citep{nicholson2017climate}.

% Spatial variation in predictive uncertainty further reflects data density and environmental heterogeneity. Regions with sparse surveillance coverage or sharp ecological transitions exhibit higher uncertainty, whereas areas with dense observations and more homogeneous conditions show tighter predictive intervals. This behavior is consistent with principled uncertainty quantification, as the model appropriately expands uncertainty where information is limited while maintaining stable estimates in well-supported regions.

\section{Discussion}

We proposed MultiDeepGP, a unified Bayesian deep learning framework for modeling multivariate mixed outcomes with complex spatial dependence. Classical multivariate geostatistical models rely on explicitly specified cross-covariance functions to ensure positive definiteness, often through constructions such as the linear model of coregionalization. Although theoretically well established, these approaches become restrictive in high-dimensional settings, require careful parametric specification, and typically rely on stationarity or separability assumptions. The difficulty is further compounded for mixed-type responses, where constructing valid and flexible cross-dependence structures becomes substantially more challenging.

By introducing a shared latent spatial representation learned through a deep Gaussian process structure, the proposed model avoids explicit cross-covariance specification and instead learns dependence implicitly through shared latent layers. This representation enables the model to jointly capture nonlinear covariate effects, cross-outcome dependence, and spatial heterogeneity within a coherent probabilistic framework, while remaining flexible in nonstationary and high-dimensional settings. Simulation results support these advantages, showing improved predictive accuracy and more reliable uncertainty quantification, particularly in nonstationary and mixed-outcome settings. The real data analysis further demonstrates the model’s ability to recover shared spatial signals that are difficult to capture using separable or stationary covariance-based approaches.

Several extensions warrant further investigation. First, the current framework assumes a fully shared latent spatial representation across all outcomes. In applications where subsets of responses exhibit stronger mutual dependence than others, a partially shared latent structure may provide additional flexibility. For example, outcomes could share a global latent component while also admitting group-specific latent processes that capture residual dependence within subsets. Such an extension would preserve the core modeling strategy of learning shared representations, while allowing heterogeneous cross-outcome dependence to be represented more explicitly. Second, the present formulation focuses exclusively on spatial dependence. In many scientific applications, mixed outcomes evolve jointly over both space and time, and their cross-dependence may change dynamically. Extending the shared latent representation to a spatio-temporal setting would allow the model to capture how outcomes co-evolve across space and time, potentially accommodating nonstationary and time-varying dependence structures. Developing scalable Bayesian deep architectures for such spatio-temporal mixed-outcome modeling represents a promising direction for future research.

Overall, MultiDeepGP provides a flexible and practically implementable framework for jointly modeling multivariate mixed spatial outcomes. By leveraging a shared deep Gaussian process representation, the approach accommodates complex nonlinear dependence across outcomes while enabling coherent uncertainty quantification within a Bayesian framework. This combination of representational flexibility and probabilistic inference makes it well suited for analyzing spatially dependent mixed-type responses in modern high-dimensional applications.

\bibliography{reference}
\end{document}